\documentclass[twocolumn]{aastex62}
\usepackage{amsmath,amstext}
\usepackage[T1]{fontenc}
\usepackage{apjfonts} 
\usepackage[figure,figure*]{hypcap}
\usepackage[normalem]{ulem}             % for strikethrough \sout{}

\usepackage{graphicx}

\renewcommand{\ion}[2]{#1\,{\sc #2}}

\shorttitle{CHIANTI Version 9}
\shortauthors{Dere et al.}

\begin{document}

\title{CHIANTI -- an atomic database for emission lines - Paper XV:  Version 9, improvements for the X-ray satellite lines}

\author[0000-0003-1628-6261]{K. P. Dere }
\affiliation{College of Science, George Mason University, 4400 University Drive, Fairfax, VA 22030, USA}

\author[0000-0002-4125-0204]{G. Del Zanna}
\affiliation{DAMTP, Center for Mathematical Sciences, University of Cambridge, Wilberforce Road, Cambridge, CB3 0WA, UK}

\author[0000-0001-9034-2925]{P. R. Young}
\affiliation{NASA Goddard Space Flight Center, Code 671,
  Greenbelt, MD 20771, USA}
\affiliation{Northumbria University, Newcastle Upon Tyne NE1 8ST, UK}

\author[0000-0002-9325-9884]{E. Landi}
\affiliation{Department of Climate, Space Sciences and Engineering, University of Michigan, Ann Arbor, MI, 48109}

\author[0000-0002-6620-7421]{R. S. Sutherland}
\affiliation{Research School of Astronomy and Astrophysics, Australian National University, Cotter Road, Weston Creek, ACT 2611, Australia}

\correspondingauthor{K. P. Dere}
\email{kdere@gmu.edu}

\begin{abstract}

CHIANTI contains a large quantity of atomic data for the analysis of astrophysical spectra.  Programs are available in IDL and Python to perform calculation of the expected emergent spectrum from these sources.  The database includes atomic energy levels, wavelengths, radiative transition probabilities, rate coefficients for collisional excitation, ionization, and recombination, as well as data to calculate free--free, free--bound, and two-photon continuum emission.  In Version 9, we improve the modelling of the satellite lines at X-ray wavelengths by explicitly including autoionization and dielectronic recombination processes in the calculation of level populations for select members of the lithium isoelectronic sequence and \ion{Fe}{xviii--xxiii}.  In addition, existing datasets are updated, new ions added and new total recombination rates for several Fe ions are included.
All data and IDL programs are freely available at http://www.chiantidatabase.org or through SolarSoft and the Python code ChiantiPy is also freely available at https://github.com/chianti-atomic/ChiantiPy.
\end{abstract}


\section{Introduction} \label{sec:intro}

The goal of the CHIANTI atomic database for astrophysical spectroscopy is to provide the atomic parameters needed to analyze spectra from high-temperature, low-density, optically-thin sources.  The main goal of Version 9 of CHIANTI is to improve the modelling of the satellite lines, by explicitly including autoionization and dielectronic recombination processes in the level populations equations for the ion. 
Previously a separate ion model (referred to as the "d" model) that  included dielectronic capture, autoionization and radiative decay was used to calculate the resulting dielectronic emissivities and these were combined with the standard model.  The new method establishes a framework to take into account density-dependent effects in the future for fully integrating level-resolved ionization and recombination processes into the CHIANTI atomic models.  For the latest Version 9, we begin with the lithium isoelectronic sequence and a set of Fe ions (\ion{Fe}{XVIII--XXIV}).  

In addition, existing datasets are updated, new ions added and new total recombination rates for several Fe ions are included.

\subsection{Goals and methods}

The intensity of an individual line of an single ion is given by 
\begin{equation} 
I\,(i \to f) \, = \, \frac{h \nu}{4\pi} \, A(i \to f) \, N_i
\end{equation} 
where $h\nu$ is the energy of the emitted photon,  $A(i\to f)$ is the radiative decay rate from initial level {\it i} to final level {\it f}, and $N_i$ is the population density of level {\it i}.  If the energy $h\nu$ is given in erg, then the units of $I(i\to f)$ are in erg~cm$^{-3}$~s$^{-1}$~sr$^{-1}$ for a single ion.

The $N_i$ are obtained by solving the equilibrium level balance equations, which model the atomic processes populating and de-populating the levels. The atomic parameters include radiative decay rates (or $A$-values), electron and proton collisional rate coefficients, autoionization rates, and level-resolved recombination rates from the next-higher ionization state. The primary goal of this paper is to describe our calculations of autoionization rates and  dielectronic recombination rates and their roles in the CHIANTI database.

In Version 3 of CHIANTI \citep{version3}, the dielectronic lines were accounted for by a model that assumed that dielectronic recombination took place only from the ground level of the higher ionization stage.   These processes were modeled by the the dielectronic ions, such as  \textit{fe\_24d}, and atomic data for these models were kept in special directories.  The parameters were based on the calculated atomic parameters of \citet{kato_safronova}.  These calculations included the $1s2snl$ and $1s2pnl$ levels with $n = 2, 3, 4$ and 5 levels and the $n=6$ levels for Fe, Ca and S and $l \le n$.  The radiative data files contained the autoionization rates as simply a decay process.

In CHIANTI version 8 \citep{version8} the atomic models for the lithium sequence ions were updated with  the collision strengths of \citet{liang_liseq}, which were provided for transitions between all bound levels with a valence electron with $n\le 5$ and core electron excitations up to $n=4$.  The radiative recombination rates of \citet{badnell_rr} were included for the bound levels of the lithium-like ions, through $n=8$.  

For the new Version 9, our approach has been to include a more complete description of each ion, by adding the autoionizing levels to the bound levels.  For the lithium sequence, this includes bound levels $1s^2nl$ ($2 \le n \le 8$; $l\le n-1$) and autoionizing levels $1s2snl$ and $1s2pnl$ ($2 \le n \le 8$; $l\le n-1$). More details for each sequence and on the method of calculation are given below.  The version 9 ion file structure is still described by the usual CHIANTI files but with the addition of a new file {\tt .auto} containing the autoionization rates for each level.  For example, the lithium-like ion Fe XXIV now contains an additional file {\tt fe\_24.auto} containing the autoionization rates. The direct inner-shell excitation and the radiative values to the bound states have been added (whenever they were not present) to the existing rate files.  Dielectronic recombination rates into the levels above the ionization potential  have been  calculated following \citet{burgess_dr} or, equivalently, equation 15 of \citet{badnell_dr}.

Once the new ion models are created, the \textit{dielectronic} ion models such as \textit{fe\_24d} are removed from the database.

\subsection{Atomic data for the improved models}

New energy levels, radiative decay rates and autoionization rates have been calculated with the AUTOSTRUCTURE (AS) code \citep{badnell_as} for the present work. Unless otherwise stated, all new radiative decay and autoionization rates have been taken from these calculations.  A drawback with AS lies in the accuracy of the calculated energy levels above the ionization potential (IP) since the autoionization and radiative rates depend on the transition energy. However, AS does include a mechanism, using a {\tt SHFTIC} file to correct the energies produced by AS.  The values used in the {\tt SHFTIC} were determined from observed energies and wavelengths and the corrected energies, the A-values and autoionization rates for all of the levels were calculated by one of us (KPD).  These values are stored in the {\tt .wgfa} and {\tt .auto} files.

Various sources for accurate level energies, particularly above the IP, exist.  We have used the NIST \citep{NIST_ASD} energy levels based on observed spectral lines where available.  Recently, \citet{rudolph} have used an EBIT electron beam trap to measure the wavelengths of photo-excited levels of highly excited species of Fe XVIII through Fe XXV illuminated with a mono-energetic synchrotron beam.  These measurements have been included for the ions we discuss in this paper.  \citet{yerokhin} report relativistic configuration interaction energy levels for the 1s2s$^2$, 1s2p$^2$ and 1s2s2p configurations of elements between argon and krypton.  These have been preferably used for the theoretical levels of these configurations for the ions for which they are available.  

The calculations of Safronova \citep{kato_safronova} are available for levels above the IP for n up to 5, and for n=6 for the ions Fe XXIV, Ca XVIII and S XVI.  More recently, these calculations for the n = 2 and 3 levels have been improved by \citet{goryaev} and these have been used to create the {\tt SHFTIC} files.  The calculations of Safronova and \citet{goryaev} are both based on a \textit{Z-expansion method} and provide wavelengths, rather than level energies.  It has been necessary to use bound state energies largely based on the NIST \citep{NIST_ASD} energies to determine level energies from the Safronova and \citet{goryaev} calculations. In the case of the lithium iso-electronic sequence, our approach in using the AS corrections in the {\tt SHFTIC} file is to use the individual level corrections for the 1s2s$^2$, 1s2p$^2$ and 1s2s2p configurations and an average correction for each of the n = 3, 4 and 5 individual configurations.  For these latter three configurations, the corrections do not vary significantly and so the n = 5 correction is then applied to the n = 6, 7 and 8 configurations.

For the ions in the lithium isoelectronic sequence it is only possible to autoionize to the ground level 1s$^2$ $^1S_0$ level of the helium-like ion.  Similarly, it is only possible to recombine dielectronically by collisions with the  helium-like ion in the same ground level.  For other ions, a larger number of levels offer the possibility of autoionization and dielectronic recombination.  For example, the beryllium-like ion can autoionize to the 1s$^2$2s and 1s$^2$2p levels of lithium-like ions.  The lithium-like ion in these levels can also dielectronically recombine to the beryllium-like ions.

\subsection{Software implementation of explicit autoionization and dielectronic recombination processes}

The principal modification to the software was to include levels of the recombining ion in the calculation of the population of the recombined ion.  In the low density approximation, only the lowest level of the recombining ion is populated and thus only one additional level is included in the solution for the level populations.  For ions where dielectronic recombination/autionization proceed {\it via} a number of levels in the recombining ion, the software includes the population of all of the levels of the recombining ion.  In this way, as the higher levels of the recombining ion become populated at finite densities, the dielectronic recombination can occur from these higher levels and alter the populations of the levels in the recombined ion giving rise to the satellite lines.  This has been previously pointed out by e.g. \citet{phillips_1983}, \citet{jacobs} and \citet{decaux_2003}.  Details about the model are given in the Appendix.

\section{New data for modeling autoionization and dielectronic recombination}

\subsection{The lithium isoelectronic sequence ions \ion{Zn}{xxviii}, \ion{Ni}{xxvi}, \ion{Fe}{xxiv}, \ion{Cr}{xxii}, \ion{Ti}{xx}, \ion{Ca}{xviii}, \ion{Ar}{xvi}}

The lithium sequence models of Version 8 have not been changed but they have been extended.  In Version 8, the bound levels with principal quantum number n through 8 and levels above the IP with n up to 4 were included.  Here, we extend the levels above the IP to n=8.  The {\it observed} bound level energies are taken from the NIST database except where noted below. The energies for the $1s2l2l'$ levels are derived from \citet{yerokhin}.  For the $1s2l3l'$ levels, the energies are derived from the calculations of \citet{goryaev}.  For the $1s2lnl'$ levels, the energies are derived from the calculations of \citet{kato_safronova} for n=4 and 5 and for n=6 where available.

These energies are used to create the AS {\tt SHFTIC} energy correction file for these ions.  The corrections derived from the observed NIST energies and the energies of \citet{yerokhin} are used directly.  For the $1s2lnl'$ levels where n = 3, 4, 5, or, 6, not all possible levels are provided by the calculations of \citet{goryaev} or \citet{kato_safronova}.  In these cases, average values of the corrections for each n level have been used.  For values of n higher than covered by \citet{kato_safronova}, the average value for the highest n level available are used.  For n = 4 and above, the corrections do not change very much with n.

\subsubsection{\ion{Fe}{xxiv}}
Energies for levels below the ionization threshold are taken from \citet{delzanna_fe_24}, and energies for several autoionizing levels were derived from the wavelengths given by \citet{rudolph}.

\subsubsection{\ion{Ti}{xx}}

Wavelengths observed by \citet{goldsmith} from the n = 3 levels have been used to extend the energy levels from the NIST database \citet{NIST_ASD}.  Further, laboratory wavelengths measured by \citet{fawcett_ridgeley} from the n = 4 levels have been included.

\subsection{The lithium isoelectronic sequence ions \ion{S}{xiv}, \ion{Si }{xii}, \ion{Al}{xi}, \ion{Mg}{x}, \ion{Ne}{viii}, \ion{O}{vi}, \ion{N}{v}, and \ion{C}{iv}}

For these members of the lithium isoelectronic sequence,  some observed energies for the bound levels are available from the the NIST database \citep{NIST_ASD}.  For the $1s2l2l'$ and $1s2l3l'$ levels, the energies are derived from the calculations of \citet{goryaev}.  For the $1s2lnl'$ levels, the energies are derived from the calculations of \citet{kato_safronova} for n=4 and 5.  For \ion{S}{xiv},  \citet{kato_safronova} also provide data for n=6.

As discussed above, the available data are used to create an optimized AS calculation of the energies, A-values and autoionization rates.  These values are stored in the usual CHIANTI  {\tt .elvlc}, {\tt .wgfa} and {\tt .auto} files.

\subsection{The beryllium isoelectronic sequence \ion{Fe}{xxiii}}

AS has been used to calculate the autoionization rates for transitions from the 1s2s$^2$2p, 1s2s2p$^2$ and 1s2p$^3$ states to the 1s$^2$2s and 1s$^2$2p states of \ion{Fe}{xxiv}. As described above a {\tt SHFTIC} correction file has been used.  The corrections for the bound levels come from the NIST database \citep{NIST_ASD}.  The energies/wavelengths of \citet{rudolph} have been used for the 1s2s$^2$2p $^3$P$_1$ and the 1s2s$^2$2p $^1$P$_1$ levels and the corrections for the remaining 1s2s$^2$2p and 1s2s2p$^2$ levels come from \citet{yerokhin_2} and \citet{rudolph}.  The previous version of CHIANTI included only the rates to the \ion{Fe}{xxv} ground level of \citet {palmeri_fe_18_25}.  

\subsection{The boron isoelectronic sequence \ion{Fe}{xxii}}

AS has been used to calculate the autoionization rates for transitions from the 1s2s$^2$2p$^2$,  1s2s2p$^3$ and 1s2p$^4$ states to the 1s$^2$2s$^2$2p$^2$, 1s$^2$2s2p$^3$ and 1s$^2$2p$4$ levels of \ion{Fe}{xxiii} and have been used to create the fe\_22.auto file.  Again, the AS calculation of the energy levels involved a {\tt SHFTIC} correction file.  The corrections were based on the wavelengths of \citet{rudolph}. The wavelengths of \citet{rudolph} have also been used for the 1s2s$^2$2p$^2$ $^2$D$_{3/2}$ and $^2$D$_{1/2}$ levels which are essentially degenerate in energy.  In the previous version of CHIANTI, only the rates to the \ion{Fe}{xxiii} ground level of \citet {palmeri_fe_18_25} were included.  

\subsection{The carbon isoelectronic sequence \ion{Fe}{xxi}}

Again, AS has been used to calculate the autoionization rates for transitions from 1s2s$^2$2p$^3$, 1s2s2p$^4$ and 1s2p$^5$ states the  to the 1s$^2$2s$^2$2p, 1s$^2$2s2p$^2$ and 1s$^2$2p$^3$ levels of \ion{Fe}{xxii} to create the fe\_21.auto file.  The wavelengths of \citet{rudolph} have been used for the 1s.2s$^2$2p$^3$ $^3$D$_1$ and $^3$S$_1$ levels of \ion{Fe}{xxii}.    In the previous version of CHIANTI, only the rates to the \ion{Fe}{xxii} ground level of \citet {palmeri_fe_18_25} were included.

\subsection{The nitrogen isoelectronic sequence \ion{Fe}{xx}}

For \ion{Fe}{xx}, the 16 autoionizing levels included are in the 1s2s$^2$2p$^4$,  1s2s2p$^5$ and 1s2p$^6$ configurations.  These levels can autoionize to the 20 levels of the 1s$^2$2s$^2$2p$^2$, 1s$^2$2s2p$^3$ 1s$^2$2p$^4$ configurations of \ion{Fe}{xxi}.  Observed wavelengths/energies of the autoionizing levels have been reported by \citet{rudolph} for the 1s2s$^2$2p$^4$ $^4$P$_{5/2}$ and for 6 other levels of the 1s2s$^2$2p$^4$ configuration by \citet{NIST_ASD}.  The remaining autoionizing levels have been optimized by using the average of the observed corrections to an unoptimized AS calculation.

\subsection{The oxygen isoelectronic sequence \ion{Fe}{xix}}

For \ion{Fe}{xix}, the 6 autoionizing levels included are in the 1s2s$^2$2p$^5$ and 1s2s2p$^6$ configurations.  These levels can autoionize to the 15 levels of the 1s$^2$2s$^2$2p$^3$, 1s$^2$2s2p$^4$ 1s$^2$2p$^5$ configurations of \ion{Fe}{xx}.  The wavelengths of the transitions from the 1s2s$^2$2p$^5$ $^3$P$_2$ and $^3$P$_1$ levels to the ground have been measured by \citet{rudolph}.  These associated energies have been used to optimize an AS  calculation for both the 1s2s$^2$2p$^5$ and the 1s2s2p$^6$ levels.  For the latter, the correction for the optimized calculation has been assumed to be the same as the measured correction for the 1s2s$^2$2p$^5$  levels.

To summarize, the energies of the 1s2s$^2$2p$^5$ and 1s2s2p$^6$ levels, the wavelengths associated with these levels and the autoionization values from these levels have been calculated with AS and updated in the CHIANTI fe\_18.elvlc, fe\_18.wgfa files and used to create the fe\_18.auto files.  Observed energies and wavelengths have been included and are given preference when used with the CHIANTI software.

\subsection{The fluorine isoelectronic sequence \ion{Fe}{xviii}}

For \ion{Fe}{xviii} we have calculated the autoionization rates from the single 1s2s$^2$2p$^6$ level using an optimized calculation of AS.  The wavelength to the ground level has been measured by \citet{rudolph} and the associated energy level has been used to optimize the AS calculation.  Autoionization rates to the  1s$^2$2s$^2$2p$^4$, 1s$^2$2s2p$^5$ and 1s2p$^6$ levels of  \ion{Fe}{xix} have been included.  

To summarize, the energy of the 1s2s$^2$2p$^6$ level and the wavelengths associated with this level have been updated in the CHIANTI fe\_18.elvlc and fe\_18.wgfa files, and the autoionization rates from this level have been placed in the new fe\_18.auto file.

\subsection{The sodium isoelectronic sequence: \ion{Al}{iii}, \ion{Si}{iv}, \ion{P}{v}, \ion{S}{vi}, \ion{Ar}{viii}, \ion{K}{ix}, \ion{Ca}{x}, \ion{Cr}{xiv}, \ion{Mn}{15}, \ion{Fe}{xvi} and \ion{Ni}{xviii}}

In Version 8 \citep{version8}, new models of the sodium isoelectronc sequence were created.  These models included 161 fine structure levels with 32 bound levels for 3s through 6h.  Levels above the IP included the 2p$^5$3s$^2$, 2p$^5$3s3p, 2p$^5$3s3d and 2p$^5$3p3d configurations.  The autoionization rates included in the Version 8 {\tt .wgfa} files have been used to create the Version 9 {\tt .auto} files.

\subsection{A summary}

To summarize, the optimized AS calculation has been used to provide the autoionization rates of the configurations mentioned above and have been used to determine the energies, A-values, and the autoionization rates included in the file of the recombined ion.  These data have been incorporated into the energy level files  {\tt .elvlc}, the radiative data file  {\tt .wgfa} the autoionization file {\tt .auto}.  For the lithium-like ions and  \ion{Fe}{xviii} and \ion{Fe}{xxii} an optimized AS calculation has been used to determine the theoretical energies of all the bound and auto-ionization levels and {it observed} levels and been inserted as well.  For the remaining Fe ions \ion{Fe}{xxi}, \ion{Fe}{xx}, \ion{Fe}{xix}, \ion{Fe}{xviii}, the AS calculation has determined the theoretical values of the levels above the IP.  The autoionization rates in the Version 8 radiative data files {\tt .wgfa} have been removed for the ions mentioned above.  For the sodium isoelectronic sequence, the Version 8 autoionization rates were transfered to a new Version 9 autoionization file {\tt .auto}.

\section{Autoionization and satellite lines}

The spectra of helium-like ions and their satellites are strong features in astrophysical X-ray spectra.  These  features were first noted by \citet{edlen_tyren}.  \citet{gabriel_jordan_hesat} suggested that the lines longward of the helium-like lines were due to transitions of the lithium-like lines, while \citet{gabriel_paget} and \citet{gabriel_1972}  provided the rate equations for predicting the intensities of these lines and a naming convention for the lithium-like satellites. 

Spectral lines created by radiative decays following dielectronic capture of free electrons into autoionizing levels play an important role in observed spectra, particularly in the X-ray wavelength region.  The lithium sequence  gives rise to many such lines as satellites to the helium sequence lines.  Often these lines are blended with  the helium sequence lines, which are important diagnostics of densities and  temperatures, so their analysis requires an understanding of the lithium sequence blends.  For example, \citet{vainshtein_safronova_78} calculated the rates needed to predict the intensities for a large number of transitions from the n=2 satellites for nuclear charge Z of 3 to 34.  Many of the n=2 lithium-like satellites are often as intense as some of the helium-like lines but many are also blended with each other and with the helium-like lines.  The exact nature of the blending varies from element to element.  For the case of the iron satellites, \cite{bely-dubau_1979} performed calculations that showed that a large number of weaker satellites from the $n=3$ to $n=16$ lithium-like levels could produce significant blending as well.
In addition, the lithium sequence dielectronic lines are useful for several spectroscopic diagnostic applications, e.g., to measure electron temperatures, departures from ionization equilibrium or thermal plasma. See, e.g., \citet{gabriel_1972} and the reviews by \citet{dubau_volonte:1980}, \citet{phillips_2008} and 
\citet{delzanna_mason:2018}.  They have often been used in solar physics studies, and recently received significant attention by the astrophysics community, after the first high-resolution X-ray astrophysical spectra obtained by the Hitomi satellite, see e.g. \citet{hitomi_perseus}.

A particularly interesting diagnostic characteristic of some of the Fe satellite lines is that their intensities are sensitive to electron density.  \citet{phillips_1983} showed that several satellite lines provide density diagnostics for densities above 10$^{13}$ cm$^{-3}$.  The density dependence is largely related to the density variation of the population of the metastable levels of the recombining ion which affects the populations of the recombined ion as a result of dielectronic recombination from levels above the ground level.  This causes some satellite lines to be stronger and some to be weaker than in the case of a low density plasma.  \citet{jacobs} and \citet{decaux_2003} have carried out a similar study for temperature and densities consistent with solar flare plasmas, magnetically-confined laboratory plasmas and laser-produced plasmas.

\subsection{Comparison with observations:  BCS flare spectra of Fe\,XXV and Fe\,XXIV, Fe\,XXIII and Fe\,XXII}

The Bent Crystal Spectrometer (BCS) \citep[BCS;][]{acton_fcs,culhane_bcs} experiment aboard the Solar Maximum Mission recorded X-ray spectra of numerous flares that produced significant \ion{Fe}{xxv} and \ion{Fe}{xxiv} emission that can be used to compare against  synthetic spectra from CHIANTI Version 9.  Recently, \citet{rapley_bcs} have pointed out some problems with the BCS data.  The main effect of the problem is that the dispersion tends to gradually increase with time during a solar flare.  This is mainly seen in the Ca spectra and occasionally in the Fe spectra.  However, this effect is quite obvious and it is possible to avoid using spectra where this occurs.  Here we have used data from the decay phase of the 1989 April 1 flare near 07:32 UT that do not appear to have the problems described by \citet{rapley_bcs}.

\begin{figure}[t]
%\epsscale{0.5}
\plotone{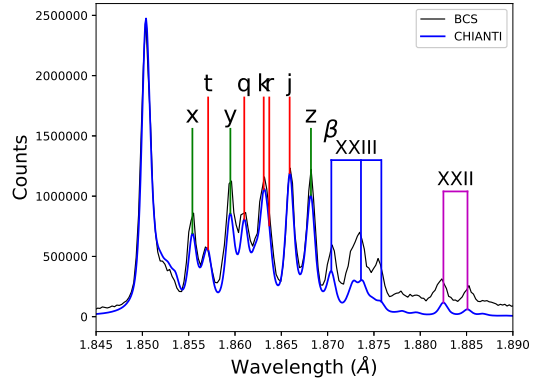}
\caption{The BCS spectra (black) of the 1989 April 1 flare during the decay phase compared with the CHIANTI 9 spectrum (blue) for a temperature of 2.12 $\times$ 10$^7$~K. Lines of \ion{Fe}{xxv} and \ion{Fe}{xxiv} are denoted with green and red vertical lines, respectively. Lines of \ion{Fe}{xxii} and \ion{Fe}{xxiii} are also indicated.}
\label{fig.bcs}
\end{figure}

Synthetic CHIANTI spectra have been calculated at a temperature of 2.12 $\times$ 10$^7$~K and are plotted in Fig.~\ref{fig.bcs} together with the BCS spectra.  A pseudo-voigt profile, available in ChiantiPy, has been used to approximate the experimental line profile.  The lines of \ion{Fe}{xxv}, aside from the resonance line r, have been noted in green and the lines of \ion{Fe}{xxiv} have been noted in red following the naming scheme of \citet{gabriel_paget}.  The temperature was chosen to give the best agreement between the \ion{Fe}{xxiv} satellite lines and  the \ion{Fe}{xxv} resonance line.  The predicted lines of \ion{Fe}{xxiii}  between 1.87 and 1.88~\AA\ are considerably weaker than observed, suggesting there is plasma at lower temperatures that is not  accounted for by the isothermal model.

\begin{figure*}[t]
%\epsscale{0.5}
\plotone{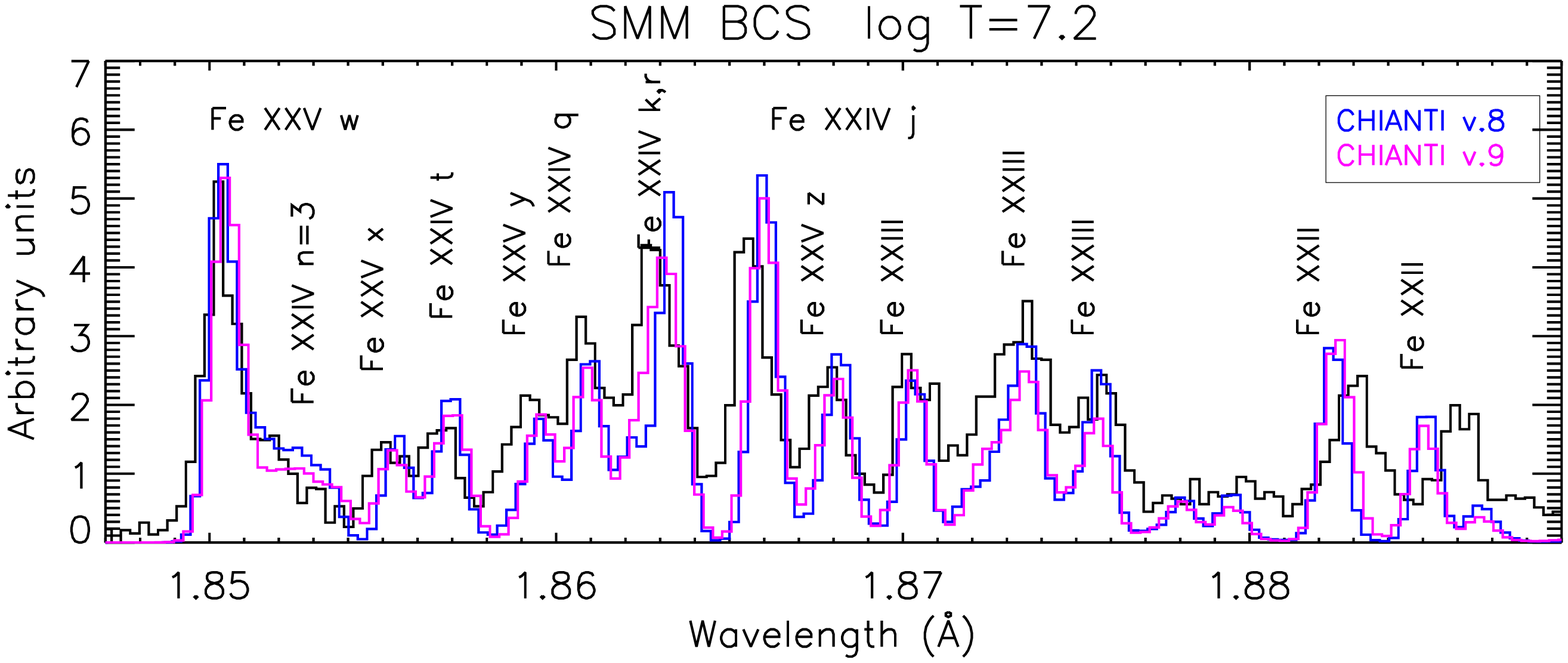}
\caption{The SMM BCS spectra (black) of the 1980 Nov 5 flare during the peak phase, as observed in the Fe XXV  channel, compared with CHIANTI v.8 and v. 9 isothermal spectra. }
\label{fig.bcs2}
\end{figure*}

\begin{figure}[!htbp]
\begin{center}
\includegraphics[angle=90, width=9cm]{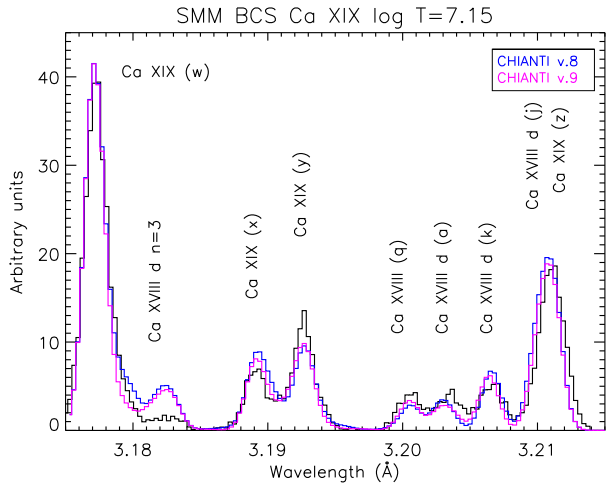}
\caption{The BCS spectra (black) of the 1980 Nov 5 flare during the peak phase, as observed in the  Ca XIX channel, compared with CHIANTI v.8 and v. 9 isothermal spectra. }
\end{center}
\label{fig.bcs3}
\end{figure}

As a further example we show  in Figures~\ref{fig.bcs2} and \ref{fig.bcs3} SMM BCS spectra (black) of the 1980 Nov 5 flare during the peak phase, as observed in the Fe XXV  and Ca XIX channels, compared with CHIANTI v.8 and v.9 isothermal spectra, calculated with the IDL codes. As it can be seen, the differences between v.9 and v.8 spectra are not large.  Note that the wavelength scale of the BCS Fe XXV channel towards the longer wavelengths (Fe XXII) is not linear, as also pointed out by \cite{rapley_bcs}.  

\section{Ionization and recombination rates}\label{sect.ionrec}

\begin{figure}[!htbp]
\plotone{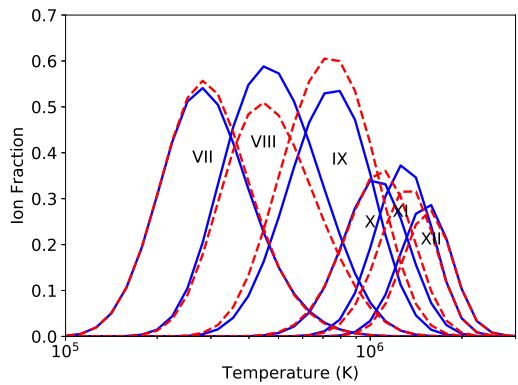}
\caption{The ionization equilibria of Fe ions VII through XII, as calculated with the current CHIANTI Version 9 rates (blue), compared to the values in version 8.0.7 (dashed red)}
\label{fig.fe_ioneq}
\end{figure}

The fits to the dielectronic recombination rate coefficients of \ion{Fe}{viii} through \ion{Fe}{xi} have been replaced.  \citet{schmidt_dr} used a combination of experimental measurements and theoretical calculations to provide the rate coefficients for \ion{Fe}{viii} forming \ion{Fe}{vii} and \ion{Fe}{ix} forming \ion{Fe}{viii}.  Using a similar approach, \citet{lestinsky_dr} provide rate coefficients for \ion{Fe}{x} forming \ion{Fe}{ix} and \ion{Fe}{xi} forming \ion{Fe}{x}.  The experimental measurements provide accurate values for the energies of the recombining ions above the IP.  Using these energies, the rate coefficients were calculated with AS.  The measurements for energies just above the IP are especially important for computing the rate coefficients at temperatures below 10$^4$ K.  The table of equilibrium ionization fractions (distributed as the file \emph{chianti.ioneq} in the database) has been recomputed for Version 9.  The ionization equilibria for the Fe ions \textsc{vii} through \textsc{xii} are shown in  Figure~\ref{fig.fe_ioneq} with values from the current version 9 and for the previous version 8.0.7. The largest change is for \ion{Fe}{viii}, for which the ion fraction is up to 30\%\ larger in CHIANTI 9.

\section{Other updated ion datasets}

\subsection{N\,I}

A mistake in the A-values for neutral nitrogen has been fixed. The nearly degenerate $^2D_{3/2,5/2}$ levels of the ground configuration  2s$^2$ 2p$^3$ were erroneously inverted.

\subsection{C\,II}

The experimental energy for the $2s^22p$ $^2P_{3/2}$ level was updated
to the value of \citet{1986ApJ...309..828C}, and the energies of the
three $2s2p^2$ $^4P_{J}$ levels were updated to the values of
\citet{2011ApJS..196...23Y}.

\subsection{Mn\,X}

The \ion{Mn}{x} model was added to CHIANTI 7 \citep{chianti7} and
featured 48 fine structure levels. All of the 38 levels of the
$3s^23p^33d$ configuration were assigned theoretical energies. Observed energies are now available from NIST for eight of the levels. These have been added to the energy level file and the wavelength file has
been updated. No other changes to the model were performed.

\subsection{Ca\,XIV}

The excitation data in CHIANTI v.8 were from \cite{landi_bhatia:2005}, and were calculated under the distorted wave (DW) approximation. It is well known that this approximation works very well 
for strong dipole-allowed transition, but typically underestimates the 
excitation of the forbidden lines, especially within the ground configuration. This was confirmed by  a recent $R$-matrix calculation by \cite{dong_etal:2012}, where the excitation rate for the  943.58~\AA\ forbidden transition is about a factor of two higher with the Dong et al.\ calculations. 
We have adopted the Dong et al.\ excitation rates, and supplemented them with A values from the recent GRASP2K calculations by \cite{wang_etal:2016}.
The ratio of the forbidden line, often observed by SoHO SUMER, with the strongest line, the resonance EUV line at 193.87~\AA\ observed with Hinode EIS, is about factor of 1.7 higher with the new model ion. This change brings into agreement SUMER and EIS observations, as shown in \cite{parenti_etal:2017}.
         
\subsection{Fe\,X}

In the analysis of the spectrum of \ion{Fe}{x} by \citet{delzanna_fe_10}, level 25 (3s$^2$ 3p$^4$ 3d $^2$D$_{3/2}$) was not assigned an observed energy or wavelength.  However, the NIST database \citep{NIST_ASD} assigns an observed energy of 511\,800 cm$^{-1}$ to this level and this has been inserted into the fe\_10.elvlc file for Version 9 and the wavelengths of the lines decaying from this level have been recalculated.

\subsection{Fe\,XII}

Energies of six levels in the \ion{Fe}{xii} model have been changed following the work of \cite{wang_etal:2018_fe_12}. The observed energies of levels 21, 38, 84 and 87 have been modified; the observed energy assigned to level 85 has been removed; and an observed energy has been assigned to level 50. No other changes to the \ion{Fe}{xii} model have been performed.

\subsection{Fe\,XVIII}

A modification was made to the \verb|fe_18.reclvl| file. The 1--91 transition was found to have an anomalously high value at the highest tabulated temperature that was traced to the original data-set of \citet{2003ApJ...582.1241G}. The value was replaced with one extrapolated from the lower temperature points.

The effective collision strengths for the bound levels were not correctly extrapolated above 7~MK when they were processed for CHIANTI~6 \citep{chianti6}, and so the data have been re-processed for CHIANTI~9.

\subsection{Fe\,XIX}

A modification was made to the \verb|fe_19.reclvl| file. The 1--160 transition was found to have an anomalously high value at the highest tabulated temperature that was traced to the original data-set of \citet{2003ApJ...582.1241G}. The value was replaced with one extrapolated from the lower temperature points.

\subsection{Fe\,XXI}

A modification was made to the \verb|fe_21.reclvl| file. The 1--179 transition was found to jump by four orders of magnitude between two adjacent temperatures, suggesting an error in the rates for this transition. Since the calculated rates were small, the transition has been removed from the file. The 1--215 transition was found to have an anomalously high value at the highest tabulated temperature and this value has been replaced with one extrapolated from the lower temperature points. Both of these problems were present in the original data-set of \citet{2003ApJ...582.1241G}.

\section{New ions: K\,IV}

The CHIANTI model consists of the 5 fine structure levels in the ground configuration.  Observed energies have been taken from version 5.3 of the NIST database \citep{NIST_ASD}.  Theoretical energies were taken from the calculations of \citet{froese_fischer_2006}.  The set of A-values come from \citet{biemont_hanson_1989}

The atomic data for \ion{K}{iv} is rather sparse.  The collision strengths have been taken from calculations of \citet{galavis}.  They provide fine-structure collisions strengths among the $^3$P level but only LS collision strengths to the $^1$D and $^1$S levels in the ground configuration.  For these latter levels, fine-structure collision strengths have been developed from these calculations by scaling the collision strengths according to the the statistical weight of the $^3$P levels.

\section{Further additions to the CHIANTI IDL programs}

\subsection{CHIANTI\_DEM}

By default, the CHIANTI\_DEM program was using the subroutine 
 \textit{xrt\_dem\_iterative2.pro} \citep{2004IAUS..223..321W}. 
 This routine, widely used in solar physics and available within SolarSoft, is based on 
the robust chi-square fitting program \textit{mpfit.pro}.  Within this subroutine, the differential emission measure (DEM) is modelled assuming a spline, with a fixed selection of the nodes.
As it turns  out, the $DEM$ solutions are quite sensitive to the choice of nodes, so
 we have replaced the \textit{xrt\_dem\_iterative2.pro}  routine with a new subroutine, called MPFIT\_DEM, where we allow for the definition of the number and location of the spline nodes.  We also simplified substantially the code and introduced the option to  input minimum and maximum limits to the spline values, which are passed to \textit{mpfit.pro}.  These limits are useful when a spectral line intensity is not measured but an upper limit can be defined. Various other options are available, as described in the header of the programs. 
The usage of the  CHIANTI\_DEM program is unchanged.

\subsection{AIA temperature responses}

\begin{figure}[!htbp]
\plotone{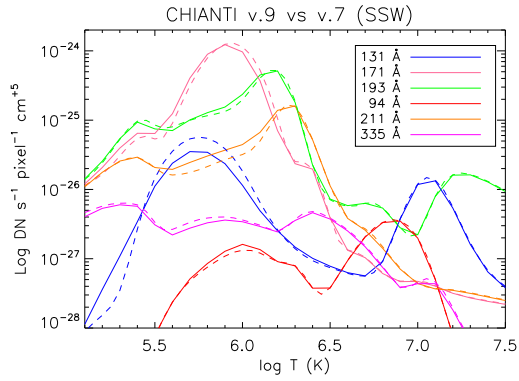}
\caption{The AIA temperature responses as calculated with the CHIANTI Version 9 ionization equilibrium and atomic data (full lines), compared to the default AIA responses available within SolarSoft (dashed lines), which were calculated with CHIANTI v.7. }
\label{fig.aia}
\end{figure}

Upon suggestion of the SDO AIA PI (Dr.~M.~Cheung), we have now added a simple IDL
program so that users can calculate the AIA temperature responses using the 
latest understanding of the decrease in the efficiency of the various broad-band filters, and the latest CHIANTI data. Users can also calculate the responses using different parameters (e.g., densities, chemical abundances) to see how they affect them. 
Figure~\ref{fig.aia} shows the AIA temperature responses as calculated with the CHIANTI Version 9 ionization equilibrium and atomic data, compared to the default AIA responses available within SolarSoft, which were calculated with CHIANTI v.7. The parameters (chemical abundances and constant pressure) are the same for both. The differences in the 131 and 171~\AA\ channel responses mainly arise due to the changes in the ionization fractions of \ion{Fe}{viii} and \ion{Fe}{ix} (Sect.~\ref{sect.ionrec}). The other differences are mainly due to updates of the ion data-sets. Note that the responses are calculated in the 25--900~\AA\ wavelength region, and there are some significant off-band contributions above 400~\AA, which were introduced in 2013 within the AIA 
programs.

\section{Conclusions}

We have improved the modelling of the satellite lines in at X-ray wavelengths with the explicit inclusion of autoionization and dielectronic recombination processes in the calculation of level populations.
The changes, compared to the previous CHIANTI versions, are not 
large, which shows that the approximate treatment of the satellite lines was relatively accurate. The explicit inclusion of these rates provides a framework for further investigations of these processes at higher densities.
Considering the further improvement in the wavelengths of the satellite lines, we believe that the present version can reliably be used for any astrophysical applications involving the satellite lines. 

All data and IDL programs are freely available online\footnote{http://www.chiantidatabase.org} and through the \emph{SolarSoft} IDL library\footnote{http://www.lmsal.com/solarsoft/}. ChiantiPy, the Python application for CHIANTI, is also freely available through \emph{Github}\footnote{https://github.com/chianti-atomic/ChiantiPy}.

\acknowledgments
GDZ acknowledges support from STFC (UK) via the consolidated grant 
to the atomic astrophysics group at DAMTP, University of Cambridge.
PRY, KPD and EL acknowledge support from NASA grant NNX15AF25G.

We would like to thank the many colleagues for their comments which helped us to uncover errors in the database and improve it. In particular, we thank John Raymond.

%\facility{facility ID}
%\facilities{facility ID, facility ID, facility ID} 
\software{ AUTOSTRUCTURE \citep{badnell_as}, Numpy \citep{numpy}, Scipy \citep{scipy}, Matplotlib \citep{Hunter:2007}, ChiantiPy, CHIANTI-IDL}

\bibliographystyle{aasjournal}
\bibliography{references}

\appendix

\section{Implementation of the new collisional-radiative method within the IDL software}

The main processes that form the doubly excited (autoionizing) states $s$ are inner-shell 
excitation of one electron in the {\it lower}  ionization stage $Z^{+r}$, and the dielectronic 
capture (DC) of a free electron by the {\it higher} (ionization stage) ion $Z^{r+1}$ in the 
state $k$. The lower ion $Z^{+r}$ in the {\it autoionizing} state $s$ can then autoionize 
(releasing a free electron) to any of the states $k$ of the  ion $Z^{r+1}$, or produce a radiative transition into any  bound state $f$ of the recombined ion. The intensity of the 
satellite line, resulting from the decay to a final bound level $f$ of the lower ion $Z^{+r}$, 
is  
\begin{equation} 
I_{sf} = N_s A_{sf}
\end{equation} 
where $A_{sf}$ is the radiative decay rate from level $s$ to the final level $f$, and $N_s$ 
is the population of the autoionizing state.

In previous versions of CHIANTI, and for the ions without autoionizing states, the level 
populations $p$ are obtained by solving the rate equations:

\begin{equation}
(A + N_{\rm e} \, (C^e + C^p) + P) \, p = b \;,
\end{equation}
where $b$ is a vector set to zeros except for the first element which is 1. The most 
important matrices are those for the spontaneous decay processes ($A$), and for the 
collisional excitation/de-excitation due to electron impact ($C^e$). Additional matrices 
for photo-excitation ($P$) and proton excitation ($C^p$) and their de-excitation processes 
are also included. 

The population of the autoionizing state due to dielectronic capture 
involves the solution of rate equations where both the recombined and recombining ions 
are included.  Thus, in CHIANTI V.9 we have modified the IDL codes in order to solve for the level populations of 
all the levels of the lower (recombined) and higher (recombining) ion simultaneously. 
In order to do this, we have extended the matrices $A, C^e, C^p$ and $P$ to include all 
the levels of the lower ion (including the autoionizing levels) and the bound levels of 
the higher ion. This means that the rates connecting the bound levels within each of the 
ions are included in the same way as in the previous versions, but we now have included
the rates for the autoionizing levels and those connecting the two ions. 
This framework naturally takes into account density effects on the satellite lines related to the populations of the metastable levels in the recombining ions. 

The population of the autoionizing state due to inner-shell excitation is calculated in a 
similar way as for the bound levels, including the inner-shell impact excitation rate
$N_{\rm e} \, C^e_{is}$ in the rate equation. The rate coefficient $C^e_{is}$ is retrieved 
from the scaled effective collision strengths stored in the {\it .scups} file.
The de-excitation rate due to a collision by a free electron is included as in the case of 
the bound levels, but this term is usually negligible.

The decay rate due to autoionization, $A^{\rm auto}_{sk}$, of the doubly-excited state $s$ 
to the state $k$ is included in the same way as the matrix $A$. The autoionization rates are 
stored in the {\it .auto} file. Note that in the case of the satellites of He-like ions, all
autoionizations go to the ground state, where most of the population is. For the recombining 
ions with metastable levels, dielectronic capture from populated levels can occur 
and thus it is included, together with autoionization rates to those levels.

Dielectronic capture is effectively a population process for the autoionizing states $s$, 
proportional to the free electron density $N_{\rm e}$ and the population $N_{\rm k}$ of the 
recombining ion in its state $k$ involved in the capture. We have therefore included this 
populating process between states $k$ and $s$ as in the matrix $C^e$ of collisional excitation. 
The rate coefficient $C^{\rm dc}_{ks}$ for the capture of the free electron by the ion $Z^{r+1}$ 
in the state $k$ into a doubly-excited state $s$ of the lower ion $Z^{+r}$ is obtained from the 
autoionization rate applying the principle of detailed balance (e.g. \cite{phillips_2008}):

\begin{equation}
	C^{\rm dc}_{ks} = \frac{h^3}{(2\pi m kT)^{3/2} } \frac{g_s}{2 g_k} 
	A^{\rm auto}_{sk} \exp \left( - \frac{ E_s - E_k} {kT } \right) 
    \label{Eq:Cap_auto}
\end{equation}
This is valid as long as the electrons have a Maxwellian distribution.
To relate the populations of the lower and higher ions we also need to include 
ionization and recombination rates. 
We recall that currently CHIANTI includes collisional ionization (CI) rates between the ground states of the ions calculated by \citet{dere_ioniz}, and total radiative recombination (RR)
and dielectronic recombination (DR) rates, also between the ground states of the ions.
The recombination rates are mainly those calculated by N.R.Badnell and colleagues. The total DR rates from the ground state of the recombining ion were obtained by Badnell by summing up the contributions of all the autoionizing states. 
The  total DR rate needs a correction, to avoid double counting the DR rates.
As we have now included autoionizing levels in the model, for consistency we need to 
calculate the total DR due to the levels included in the model and originating from the 
ground level of the higher ion,  and subtract this quantity  from the total DR rate.
The remaining rate is  added in the matrix as a term connecting the two ground states.

The calculation of the total DR rate due to the autoionizing levels included in the model is not trivial, as in principle each 
autoionizing state could decay to another autoionizing level, as well 
as decay to a bound level $f'$ of the recombined ion, or autoionize to a level
of the recombining ion. We neglect the first process for two reasons.
First, we note that the other two are the main ones, 
although we note that  in some cases we do have some autoionizing levels that mainly decay to other autoionizing levels. The second reason  is that the 
total DR rates have been calculated neglecting this cascading process. 
The total rate is therefore calculated as 
\begin{equation} 
 c\;  \sum_s  \sum_k  \frac{g_s}{ 2 g_k} \;	A^{\rm auto}_{sk} \;
 e^{ - \frac{ E_s - E_k}{ kT }}\;
\frac{\sum_{f'<s} A_{sf'} }{  \sum_k A^{auto}_{sk} + \sum_{f'<s} A_{sf'} }
\end{equation} 
where the constant $c= h^3  \; (2\pi m kT)^{-3/2} $

The CI rates are included as they are available in CHIANTI, i.e., connecting the ground states. 
The RR rates are now included in two different ways. For most ions, the total RR rates from the ground state of the recombining ion are included in the matrix to connect to the ground state of the recombined ion.

For some ions, we have now introduced the level-resolved RR rates as calculated by N. Badnell. They are included with a new file, with the extension 
{\tt .rrlvl}, with a format similar to that of the previous {\tt .reclvl} files. 
We have included these level-resolved rates into the matrix. They typically increase the populations of the lower levels by 10\% or so, but for higher levels can be the only populating process. 
To avoid double counting, as in the case of the DR rates, we sum the total RR of the level-resolved rates and subtract this value from the total RR. Any residual total RR is added as a rate connecting the two ground states.

Contributions to level populations due to cascades from bound levels is therefore now naturally included in the model ions, although for many ions most of the RR occurs into high-lying levels that are not currently included in the model. For many ions we have therefore retained the corrections due to 
these cascading effects as in the previous CHIANTI versions.
Prior to CHIANTI 9 the radiative recombination rates were implemented
as a post-processing step once the level population equations of the standard
CHIANTI model had been solved, as described in \citet{chianti5}.
These RR  rates are stored in the {\tt .reclvl} files.
For the important \ion{Fe}{xvii--xxiii} ions, the rates 
are from the calculation of \citet{2003ApJ...582.1241G} and the rate
into a level includes both the direct radiative recombination rate and
the indirect recombinations that come from cascading from higher
levels, \emph{including} the autoionization levels populated by
dielectronic capture.
The effects of these rates on the level populations is approximated with 
a correction, after the matrix is inverted, as in previous  CHIANTI versions.

Note that a model CHIANTI ion can only have either  direct level-resolved RR 
rates as in the {\tt .rrlvl} files or the level-resolved RR including cascades as in 
the {\tt .reclvl} files. Also note that the {\tt .reclvl} files can only have recombination
from the ground state, while the {\tt .rrlvl} files can in principle include recombination from excited states. Finally, note that the CHIANTI programs assume that all the rates in these two files are on the same temperature grid, although the files in principle could have different temperature grids for each transitions.

After the matrices are populated and the populations obtained,
we then normalise the populations of the lower ion so the total is one, as 
in the case of ions without autoionizing states.
Note that the relative population of the two ions as obtained 
solving the two-ion rate equations can sometimes be different than what is obtained by assuming that the ion populations are all in the ground state (which is what is used in CHIANTI to calculate the relative ion charge state distributions).
However, this effect is small and is not considered in this version 9.

Finally, we note that the new v.9 IDL codes are compatible with earlier 
CHIANTI v.8 data files, but  the earlier CHIANTI v.8 IDL programs should not be used with the new v.9 format files. 

%------------------

\end{document}